\documentclass[11pt]{article} 
\usepackage{graphics}
\usepackage{latexsym}  
\newcommand{\mathR}{{\rm I\! R}}
\newcommand {\be}{\begin{equation}} 
\newcommand {\ee}{\end{equation}}
\newcommand {\bea}{\begin{eqnarray}} \newcommand
{\eea}{\end{eqnarray}} \newtheorem{theorem}{Theorem}
\newtheorem{conjecture}{Conjecture}
 \newtheorem{lemma}{Lemma}

\title{A handlebody calculus for topology change}
\author{H.F.Dowker${}^{a}$ and R.S.Garcia${}^{b}$\\
        $\;$ \\
        Blackett Laboratory,\\
         Imperial College of Science Technology and Medicine,\\
        London SW7 2BZ, United Kingdom.}
\date{5 November 1997}

\begin{document}
\begin{titlepage}
\maketitle
\begin{abstract} 
\thispagestyle{empty}
We consider certain interesting processes in quantum gravity 
which involve a change of spatial topology. 
We use Morse theory and the machinery of 
handlebodies to characterise topology changes as suggested 
by Sorkin. Our results support the view that 
that the pair production of Kaluza-Klein
monopoles and the nucleation of various higher dimensional objects 
are allowed transitions with non-zero amplitude. 
 
\end{abstract}
\vspace{4.cm}
\noindent ${}^a$ {\small dowker@ic.ac.uk}\\
\noindent ${}^b$ {\small garciars@ic.ac.uk}
\end{titlepage}

\section {Introduction} \label{introduction}
The question of whether the topology of space can change is a basic
one in the search for a theory of quantum gravity. The theorems of
Geroch \cite{geroch67} and Tipler \cite{tipler} are widely understood to
show that there is no topology change in classical general relativity,
so that we should look to the quantum theory to see it, if it
occurs. Though the definitive statement about the occurrence of
topology change may well have to wait until we have a fully developed theory
of quantum gravity it is nevertheless generally believed that topology
change {\it does} happen.  A general calculus for topology change
within the Sum Over Histories (SOH) approach, based on Morse theory,
has been suggested by Sorkin \cite{sorkin90}. In this paper we will review this
picture, and use it to investigate certain interesting topology
changing processes. In the rest of the introduction we set up some
terminology and outline our aims and results.
         Let an $n$-geometry $(M,g)$ consist of an $n$-dimensional
         manifold $M$ and a metric $g$ on $M$ --strictly, a geometry is
         an equivalence class of such pairs under diffeomorphism.  A
         {\it topology change} in $n$ spacetime dimensions is a
         transition from a Riemannian ($n-1$)-geometry $(W_0,h_0)$ to
         another Riemannian ($n-1$)-geometry $(W_1,h_1)$ in which
         $W_0$ and $W_1$ are non-diffeomorphic.

We call $(M, V_0, V_1)$ a {\it smooth manifold triad} if $M$ is a
compact smooth $n$-\nolinebreak manifold whose boundary is the disjoint union of
the two closed submanifolds $V_0$ and $V_1$, $\partial M = V_0 \uplus
V_1$.  Given two closed smooth $(n-1)$-manifolds, $W_0$ and $W_1$, a
{\it topological cobordism} from $W_0$ to $W_1$ is a 5-tuple $(M, V_0,
V_1, d_0, d_1)$ where $(M, V_0, V_1)$ is a smooth manifold triad and
$d_i: V_i\rightarrow W_i$ is a diffeomorphism, $i = 0,1$.  Cobordism
gives rise to an equivalence relation on the set of $(n-1)$
manifolds. We say that $W_0$ and $W_1$ are in the same cobordism class
if a topological cobordism exists between them. A {\it Lorentzian
cobordism} from geometry 
$(W_0, h_0)$ to $(W_1, h_1)$ is a 6-tuple $(M, V_0,
V_1, d_0, d_1, g)$ where $(M, V_0, V_1, d_0, d_1)$ is a topological
cobordism and $g$ is a Lorentzian metric on $M$ such that
$(d_i^{-1})^*(g_{|_{V_i}}) = h_i$, $i = 0,1$ and such that $V_0$ is a
past spacelike 
boundary and $V_1$ is a future spacelike boundary. We will often drop the
explicit mention of the diffeomorphisms in what follows and unless
otherwise stated all cobordisms $M$ will be compact.

        A necessary and sufficient condition for a topological
        cobordism to exist between a given pair of manifolds is
        that their Steifel-Whitney and 
Pontrjagin numbers coincide when both are
        oriented or just their Steifel-Whitney numbers in the non-oriented case
        \cite{milnor74,stong68}. Hence the number of cobordism classes
        equals that of distinct combinations of Stiefel-Whitney and Pontrjagin
        numbers, which has a finite value, depending on the
        dimension. As it happens, all $3$-manifolds are cobordant, while
        $4$-manifolds divide into four cobordism classes.

Now that we have explained what we mean by topology changing
transitions between two spacelike hypersurfaces, we must decide how to
investigate them. Among the different approaches to quantum gravity,
the Sum Over Histories affords the most natural expression for
topology changing transition amplitudes:

\begin{equation}
\langle W_1, h_1; W_0,h_0 \rangle = \sum_{(M, V_0, V_1, d_0, d_1)} 
\omega(M,
d_0, d_1) \;\; \int_{\cal C} Dg e^{iS[g]}
\end{equation}

\noindent 
where the sum is over topological cobordisms and ${\cal C}$ is a class
of metrics, $g$, on $M$ such that $(d_i^{-1})^*(g{|_{V_i}})=h_i$, $i =
0,1$. The weight $\omega(M, d_0, d_1)$ will not concern us here but
is discussed in \cite{sorkin85}.  Although this formal
expression is far from being defined, and indeed may never be so
without recourse to a possibly discrete underlying theory,
we can already draw some
conclusions from its general form. For example, if $W_0$ and $W_1$ are
not cobordant then the amplitude for the topology change is zero.

There are various proposals for the type of metrics over which the
functional integral runs for each topological cobordism $M$. Following
Sorkin \cite{sorkin97} we start
with the view that the integral should be over all Lorentzian metrics
but this immediately raises a problem. In the event of topology
change, the geometry $(M,g)$ cannot be both Lorentzian and causally
ordered. This follows from the following theorem of Geroch
\cite{geroch67}:
        
        \begin{theorem}[Geroch,1967]\label{gerochT} 
If a smooth triad $(M, V_0, V_1)$, with $V_0$ and $V_1$ closed, admits
a time-orientable Lorentzian metric $g$ without closed timelike curves and
such that $V_0$ and $V_1$ are spacelike with respect to $g$, then $V_0
\cong V_1$ and $M\cong V_0\times I$ where $I$ is the 
unit interval, {\it i.e.}, there is no topology change.
\end{theorem}

So which do we choose to keep: causal order or the equivalence
principle?  Following Sorkin \cite{sorkin86a} we plump for casual
order. For one thing, if we were instead to insist on globally
time-orientable 
Lorentzian metrics this would rule out the production of Kaluza-Klein
monopole-antimonopole pairs since there does not exist such a metric
on any topological cobordism for this process \cite{sorkin86a,sorkin86b}. Also, if causal
sets are the correct description of the discrete substructure of
spacetime then causal order is more fundamental than metric
\cite{sorkin97}. Pursuing this route, however, means we must allow
singularities of some sort in the geometries $(M,g)$ that contribute
to the amplitude for a topology changing process. So what
singularities are allowed?
Sorkin has suggested that Morse theory (see {\it e.g.}
\cite{milnor63}) furnishes the
appropriate metrics that are Lorentzian {\it almost} everywhere and
exist on all topological cobordisms.

A {\it Morse function} on a manifold $M$ is a smooth function $f: M\nolinebreak 
\rightarrow\nolinebreak\mathR$ such that $\partial_\mu f$ vanishes only at a finite
number of points ${p_k}$ where the Hessian $\mbox{$\partial_\mu
\partial_\nu f$} _{|p_k}$ is a non degenerate matrix. The {\it Morse
index} $\lambda _k$ of each critical point $p_k$ is the number of
negative eigenvalues of the Hessian matrix evaluated at $p_k$. The critical values of $f$ are the values it takes at the critical points; 
we will often denote them $c_k = f(p_k)$. The
abundance of Morse functions on a manifold is enough to ensure the
following \cite{dubrovin95, milnor65}

        \begin{theorem}\label{morsexistT} For any smooth triad, 
$(M,V_0,V_1)$, there exists a Morse function $f: M\rightarrow [0,1]$ such that:
\begin{enumerate}
\item $f^{-1}(0) = V_0$ and $f^{-1}(1) = V_1$ \item $f$ has no critical
points on $\partial M = V_0 \uplus V_1$
\end{enumerate}
\end{theorem}

Then given any Riemannian metric $G$ on $M$ and a real number $\zeta>
1$, we can construct an almost Lorentzian metric $g$ associated with
$f$ as follows:
\begin{equation}
\label{eq.morse.g}
g_{\mu\nu} = \partial _\rho f\partial _\sigma f G^{\rho\sigma}
G_{\mu\nu} - \zeta \partial _\mu f\partial _\nu f
\end{equation}
and we call this a Morse metric. It is Lorentzian everywhere except
for the Morse points and $G^{\mu\nu}\partial_\nu f$ defines a timelike
direction.  If moreover Riemannian metrics are given on $V_0$ and
$V_1$, we can demand that $g$ has the correct restrictions by choosing
$G$ appropriately.  We summarise these statements \nolinebreak as

\begin{lemma}\label{lemmametric} Let $(M, V_0, V_1, d_0, d_1)$ be a 
topological cobordism between $W_0$ and $W_1$ and let $h_0$ and $h_1$
be Riemannian metrics on $W_0$ and $W_1$. Then there exists a Morse metric
$g$ on $M$ such that 
                $(d_i^{-1})^*(g_{\mid _{V_i}}) = h_i$, $i=0,1$.

\end{lemma}

The proof is given in Appendix B. Any topological cobordism has an infinite number of Morse metrics
associated with it.  We call each such geometry $(M,g)$ an Almost
Lorentzian (AL) cobordism and restrict the functional integral to be
over such cobordisms. Since these geometries are singular, it will be
necessary to extend the definition of the action $S$ to these
cases \cite{louko97}. 
Note that in this view the critical points are not to
be sent to infinity as in \cite{alty95,yodzis73} but rather remain part
of the spacetime and indeed the causal order is well defined with the
Morse points present.

Now, Sorkin suggested that it might be necessary to impose a 
stronger condition on the set of contributing metrics.  This
observation is motivated by a very simple example in $(1+1)$
dimensions: quantum field theory on the trousers cobordism. 
The $(1+1)$-dimensional trousers admits an  everywhere flat AL metric 
        with a single index one Morse point at the
        crotch which singularity is the source for an infinite burst
        of energy of a scalar quantum field propagating on the
        trousers \cite{anderson86,manogue88}.  Anderson and DeWitt
        have argued that this provides evidence against topology
        change. But the regular propagation of a quantum field 
        on the $(1+1)$ ``yarmulke'' topology, a hemisphere
        mediating the transition $\emptyset
        \rightarrow S^1$, suggests that it might be a particular
        feature of the trousers topology, and not a general flaw of
        all nontrivial cobordisms, that causes the unphysical energy
       burst \cite{daughton}. A crucial difference between the trousers and the
        yarmulke topologies is that the former has a causal
        discontinuity whereas the latter does not
(roughly speaking a 
causal discontinuity is a
        discontinuous change in the volume of the causal
        past or future of a continuously varied point\cite{sachs73}). 
Generalising this idea Sorkin conceived the
        following conjectures:
\begin{conjecture}
A quantum field propagating on an AL cobordism $(M,g)$ has an
unphysically singular behaviour if and only if $(M,g)$ is causally
discontinuous.
\end{conjecture}
\begin{conjecture} An $n$-dimensional AL cobordism $(M,g)$ is
causally discontinuous if and only if the Morse function from which
$g$ is constructed has either an index $1$ or index $(n-1)$ critical
point \cite{borde}. 
\end{conjecture}

\begin{figure}[ht]
\centering
\rotatebox{270}{\resizebox{!}{2.5in}{\includegraphics{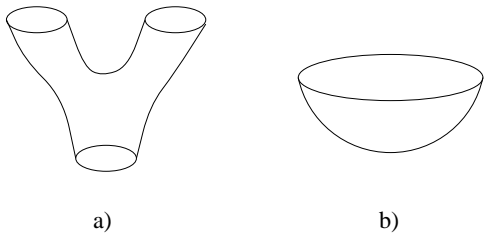}}}
\caption{{\small The $1+1$ trousers and yarmulke cobordisms.} \label{trousers}}
\end{figure}

        Plausibility arguments for the second conjecture are 
        given in \cite{borde} \cite{surya}, 
but rigorous demonstration of both is
        work in progress. Now in addition we take  the singular
        behaviour of the quantum perturbations around these causally
        discontinuous backgrounds to be an indication of 
their contribution to the SOH being infinitely suppressed. Evidence for this is
        presented in \nolinebreak \cite{louko97}, though the issue is clouded by the
        fact that the backgrounds considered there 
are not stationary points of the
        action. We can now eliminate from the SOH those topological
        cobordisms which do not admit Morse functions without index 1
        or ($n-1$) points.  The remaining AL cobordisms, the causally
        continuous ones, are our candidates to contribute to the SOH
        transition amplitude. We will denote them Causally Continuous
        AL (CCAL) cobordisms. The remainder of the article is based on
        this criterion; thus we work under the assumption that the conjectures hold.

In section \ref{handlebody} we describe how to use handlebody decompositions as a
method to identify CCAL cobordisms;\footnote{We thank Sumati Surya for
  suggesting handlebodies as a technique.} in particular we prove an identity
that will allow us to deduce handlebody decompositions in high
dimensions from lower dimensional ones. In section \ref{instantons} 
we apply this
technique to analyse various interesting topology changing processes
of semi-classical decay in quantum gravity. We find a favourable
handlebody decomposition for their respective instanton cobordisms,
and hence verify that these processes can occur.  As a preliminary we 
demonstrate that if the topological non-trivialities of a non-compact 
cobordism $L$ are confined to some compact region $M\subset L$ then the 
Morse metrics on $M$ give rise to Morse metrics on $L$ with 
the same critical structure. In the last section we 
summarise our results and list some open problems.

\section{Handlebody decompositions}\label{handlebody}

Define the {\it Morse structure} of a Morse function $f$ on $M$ to be
a complete ordered list $\{(p_k, \lambda _k): k = 1,\dots r\}$ 
of its Morse points and
corresponding Morse indices.
As we shall see, a handlebody decomposition of a
manifold, $M$, implies the existence of Morse functions
on $M$ with totally determined Morse structure. The following
definitions follow very closely the first pages of Kirby's book 
\nolinebreak \cite{kirby89}.  
They make extensive use of the concepts of closed or
open $n$-balls,
$n$-spheres and their respective boundaries, which are listed
here:
\[
\begin{array}{rlrl}
B^n = &\{ x \in \mathR ^n : |x| ^2 \leq 1 \} &\mbox{~~~} S^n = &\{ x \in
\mathR^{n+1} : |x| ^2 =1 \} \\ \\ \partial B^n = &S^{(n-1)} &\mbox{~~~}
\partial S^n = &\emptyset \\ \\
\dot{B}^{n} =& \{x\in \mathR ^n: |x| ^2 < 1\} & & 
\end{array}
\]  
\noindent By $\dot{A}$ we mean the interior of the set
$A$, {\it i.e.}, the largest open set contained in $A$. Note for
future reference that when $A$ is 
a subset of the manifold with boundary $M$ and $\partial M$ is not empty 
$\dot{A}$ may contain part of it.

 A {\it handlebody decomposition} of an $n$-dimensional compact
manifold $M$ is a nested sequence of manifolds 
$\emptyset=M_{-1}\subset M_0 \subset
M_1 \subset \cdots \subset M_r = M$ where $M_k$ is obtained by
adjoining a $\lambda _k$ handle to $M_{k-1}$, {\it i.e.}, $M_k = M_{k-1} +_{h_k}
B^{\lambda_k}\times B^{n-\lambda _k} $ via an embedding, $h_k : \partial 
B^{\lambda _k}\times B^{n-\lambda _k} \hookrightarrow \partial M_{k-1}$ of the 
boundary of the
$\lambda _k$ handle into the boundary of $M_{k-1}$. Note that $M_0 = B^n$ in
any such handlebody sequence. This definition involves two operations
whereby a pair of manifolds with boundary can be combined: adjunction
($+$) and product ($\times$).  In appendix \ref{m&gmanappend} we describe how endowing the adjunction or the product of two manifolds with a differentiable structure
raises the issue of smoothing corners. We here simply intone the
slogan ``corners can be smoothed'' ({\it e.g.} $B^n$ and $(B^1)^n$ are equivalent as far as we are concerned).

Associated with any smooth handlebody decomposition is a Morse
function $f: M \rightarrow [0,1]$ with as many critical points as handles being
attached.  For the 
$r+1$-handled-body in the definition,  the function $f$ would
have $r+1$ non-degenerate critical points, $\{p_k\}$, $k = 0,1,\dots r$, 
which can be taken to lie in different level surfaces, {\it i.e.}, $
f(p_0) < f(p_1)<\cdots < f(p_r)$. Each critical point may be located at the
centre $(0,0)$ of $B^{\lambda _k} \times B^{n-\lambda _k}$; then $B^{\lambda
_k} \times \{\vec{0} \} $ is the descending manifold and $\{\vec{0} \} \times 
B^{n-\lambda _k}$ the
ascending manifold.  By this we mean that around $p_k$ the function
$f$ admits an expansion (Morse lemma \nolinebreak \cite{milnor63}):
\begin{equation}
f(q) = f(p_k) - x_1^2 - x_2^2 - x_{\lambda _k}^2 + x_{\lambda _k+1}^2 + \cdots
+x_n^2
\end{equation}
The first $\lambda _k$ local coordinates parametrise $B^{\lambda _k}$, the last
$n-\lambda _k$ local coordinates parametrise $B^{n-\lambda _k}$ and $p_k$ is
identified as a Morse point of index $\lambda _k $. In other
words, we can define $f$ following the sequence of manifolds. It is 
zero at some point of $M_0$, which is the index $0$ critical point
${p_0}$; it then increases in a regular way except for the critical
point associated with each handle attachment.

The Morse function, $f$, associated with the handlebody decomposition
given above is 1 on the boundary of $M$. As such, it is
appropriate for the case of the topology change from the empty set to
$\partial M$.  We are interested in the more general case of
topology change from $V_0$ to $V_1$. In that case we have a manifold,
$M$, whose boundary is the disjoint union of $V_0$ and $V_1$.  A
generalised handlebody decomposition of $M$ is a nested sequence $V_0
\times B^1 = M_0 \subset M_1 \subset \dots M_r = M$ where $M_k$ is
obtained by attaching a $\lambda _k$ handle to $M_{k-1}$. But now there is a restriction
on each embedding $h_k$: its image must not intersect the initial 
$V_0$ component of the boundary of $M_k$. In this handlebody 
calculus, the addition of each handle can be thought of as an elementary 
topological transition from $\partial M_{k}$ to $\partial M_{k+1}$.

        A useful property of handlebody decompositions is ``right
        distributivity'' of a $B^m$. It allows us to deduce from a handlebody
        decomposition for a manifold $M$ a whole series
        of higher dimensional handlebodies for the manifolds $M \times
        B^m$.  The crucial point is that the new $B^m$ factor does not
        actively partake in the induced imbedding $\partial B^\lambda \times
        B^{n+m-\lambda } \hookrightarrow \partial (M \times B^m)$. More
        explicitly we have the following\footnote{We omit mention of
        the particular embeddings; one is understood with each
        handle in the sum (a total of r handles are attached in
        succession to the initial ball.)}:
 \begin{lemma}\label{lemmaBD} Let $M$ be an $n$-dimensional 
                manifold; then if
\[
M = B^n + \sum_{k=1}^{r} B^{\lambda _k}\times B^{n-\lambda _k}
\]
it follows that
\[
M \times B^m = B^{n+m} + \sum_{k=1}^{r} B^{\lambda _k}\times B^{n+m -\lambda _k}
\]
\end{lemma}
 {\bfseries Proof.} First we claim that if 
 $L$ is an $n$-dimensional manifold such that
\[
M = L + B^{\lambda } \times B^{n-\lambda } 
\]
then
\[
M\times B^1 = L\times B^1 + B^{\lambda }\times B^{n+1-\lambda }
\]        
\noindent Provided that the claim holds one can proceed by induction in m, the
dimension of the right-factored ball, to infer that
\begin{equation}
\label{Bmfactor}
M\times B^m = L\times B^m + B^{\lambda }\times B^{n+m-\lambda }
\end{equation}
The only subtle point about this induction process is
the diffeomorphism $B^l  \times B^1 \cong B^{l+1}$, but this
is again the problem of smoothing corners. Then by applying 
eq.(\ref{Bmfactor}) at each step of a handlebody decomposition one obtains:
\begin{eqnarray*}
M\times B^m &=& M_{r-1} \times B^m + B^{\lambda _r}\times B^{n+m-\lambda _r} \\ &=&
M_{r-2} \times B^m +B^{\lambda _{r-1}}\times B^{n+m-\lambda _{r_1}} +
 B^{\lambda _r}\times B^{n+m-\lambda _r} \\ &=& \cdots = M_j\times B^m +
 \sum_{k=j+1}^r B^{\lambda _k}
\times B^{n+m-\lambda _k} = \cdots \\ &=& B^{n+m} + \sum_{k=1}^{r}
B^{\lambda _k}\times B^{n+m-\lambda _k}
\end{eqnarray*}

        Finally we prove the initial claim. Given that $M\cong L
        +_h B^{\lambda }\times B^{n-\lambda }$ through the embedding $h: 
        \partial B^{\lambda }\times B^{n-\lambda } \hookrightarrow \partial L$, we define
\[\tilde{h}: \partial B^{\lambda } \times (B^{n-\lambda }\times B^1)\hookrightarrow 
\partial L \times B^1 \subset \partial(L \times B^1)\] 
by
\begin{displaymath}
\tilde{h}(x,t) = \left(h(x),t \right)
\end{displaymath}
where $x\in \partial B^{\lambda } \times B^{n-\lambda }$ and $t\in B^1$. This new embedding 
induces a  map $f: (L +_h B^{\lambda }\times B^{n-\lambda })\times B^1\rightarrow L\times B^1
+_{\tilde{h}} B^{\lambda }\times B^{n+1-\lambda }$ given, in the terminology of  
appendix \ref{m&gmanappend}, by: 
\be
\label{diffeo}
f(([z]_h,t)) = [(z,t)]_{\tilde{h}} \ee

\begin{figure}[ht]
\centering
\rotatebox{270}{\resizebox{!}{4.5in}{\includegraphics{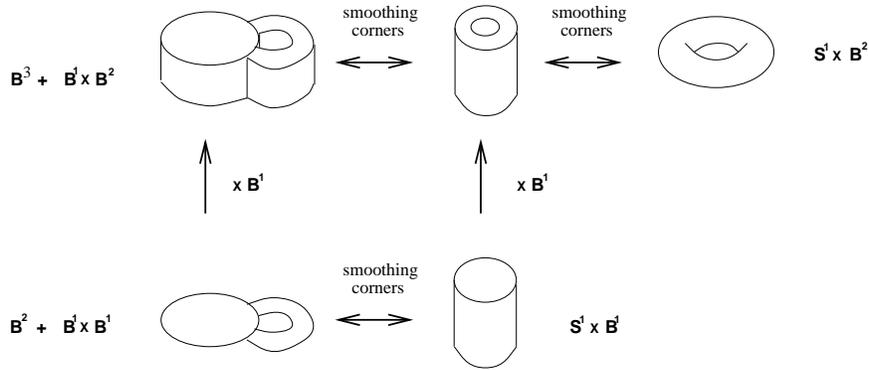}}}
\caption{\protect\parbox[t]{8.3cm}{\small Right $B^1$ distributivity
lifting a 2-dim handlebody, the hollow cylinder (bottom-right corner), to a 2-dim handlebody, the solid torus (top-right corner). } \label{bfactor}}
\end{figure}

        The map $f$ is well defined -- independent of
        class representative -- and since the same holds for its
        obvious inverse, $f$ is a bijection. It is also a
        homeomorphism of topological spaces.
        It can be shown that $f$ has differentiable local representatives even
        at the smoothed corner set \cite{kirby77} once it has been 
        composed with the relevant smoothing maps. Thus $f$ is a diffeomorphism;
it expresses distributivity between adjunction and product of manifolds with
boundary. Fig \ref{bfactor} 
        illustrates a simple case of $B^1$ distributivity: the handlebody for the annulus $S^1\times B^1 \cong B^2 + B^1 \times B^1$ gives rise
        to the handlebody $S^1\times B^2 \cong B^3 + B^1 \times B^2$. 
\\

\begin{figure}[ht]
\centering
\rotatebox{270}{\resizebox{!}{2.5in}{\includegraphics{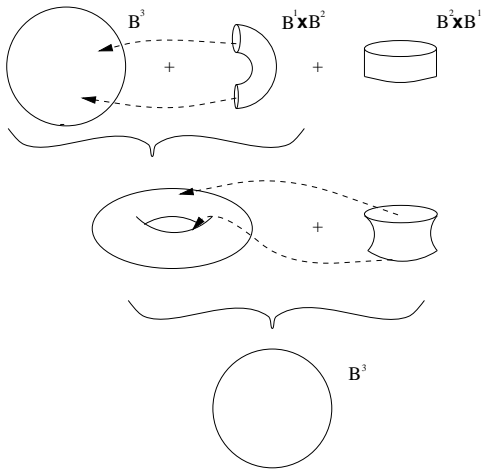}}}
\caption{{\small ``Redundant'' decomposition of the 3-ball.}\label{redundant}}
\end{figure}

With the machinery of Morse theory and handlebodies in hand we can
investigate topology changing processes. First of all the content of the
conjectures translates into the following statements.
 If a smooth triad $(M, V_0, V_1)$ has a
handlebody decomposition which does not include a $B^1\times B^{n-1}$
nor a $B^{n-1}\times B^1$ handle, then it admits a CCAL metric and
according to the premises of this paper $M$ is to be included in the
SOH for the process. On the other hand, if a smooth triad has
a handlebody decomposition which does contain a 1-handle or an
$(n-1)$-handle then we cannot draw the contrary conclusion. For
example consider two decompositions of $B^3$ (Fig \ref{redundant}):
\be
\label{longdec} 
B^3 = B^3 + B^1 \times B^2 + B^2 \times B^1 
\ee 
\be
\label{shortdec} 
B^3 = B^3 
\ee

In view of (\ref{longdec}) alone we would be wrong to
conclude that $B^3$ supports no CCAL cobordisms for the creation of
$S^2$ since (\ref{shortdec}) shows that $B^3$ does support causally
continuous cobordisms.

       However, the Morse inequalities \cite{milnor63} do furnish a
       sufficient --but not necessary-- criterion for automatically
       discarding certain cobordisms. Consider the triad
 $(M,V_0,V_1)$.  Let $\beta _\lambda (M, V_0) $ be the
       $\lambda ^{\rm th}$ Betti number of $M$ relative to $V_0$ and
       let $\mu _\lambda $ denote the number of critical points of
       index $\lambda$ of a Morse function $f: M \rightarrow [0,1]$
       with $f^{-1}(0)=V_0$ and $f^{-1}(1)=V_1$.  Then a weak version
       of the Morse inequalities establishes that:
\begin{equation}
\label{morsein}
        \mu_\lambda \geq \beta _\lambda (M,V_0) \mbox{~~~~~~}
\end{equation}
So if the first or $(n-1)^{\rm th}$ homology of $M$ relative to $V_0$
has non-trivial torsion free part, any Morse function on $M$ must have
index 1 or index $(n-1)$ points.

As an example consider the cobordism $B^4 \times S^1 $ for creation of
an\linebreak $S^3 \times S^1$ ($V_0$ is empty here).  We can compute its
homology using the Kunneth formula \cite{greenberg81} for the homology groups of the
product of two spaces when both have torsion-free homologies, namely $H_q(X \times Y)= \sum_{p=0}^{q}
H_p(X)\otimes H_{q-p}(Y)$. Applying this to $B^4 \times S^1$ gives:
\begin{eqnarray*}
H_1(B^4 \times S^1) &=& H_0(B^4) \otimes H_1(S^1) + H_1(B^4) \otimes
H_0(S^1) \\ &=& Z \otimes Z + 0 \otimes Z = Z
\end{eqnarray*}
The same, applied to $B^2 \times S^3$ gives:
\begin{eqnarray*}
H_1(B^2 \times S^3) &=& H_0(B^2) \otimes H_1(S^3) + H_1(B^2) \otimes
H_0(S^3) \\ &=& Z \otimes 0 + 0 \otimes Z = 0 
\end{eqnarray*} 
Thus $\beta_1(B^4 \times S^1) = 1$, while $\beta_1(B^2 \times S^3) =
0$. In conjunction with eq.(\ref{morsein}) $\beta_1(B^4 \times S^1) =
1$ tells us that there are no Morse functions on $B^4 \times S^1$
without index 1 critical points and so this cobordism does not admit
CCAL metrics. But for general $M$ a vanishing $\beta_1$ does not
guarantee that there is an allowed Morse function on $M$, since there
is no reason why Morse functions should exist that saturate the
inequalities \cite{surya}. In particular, from $\beta_1(B^2 \times
S^3) = 0$ alone we could not infer that $B^2 \times S^3$ admits CCAL
metrics. It is only in view of the handlebody decomposition given
earlier that we can so conclude.

\section{Handlebodies and instantons for 
semi-classical decay in quantum gravity}\label{instantons}

Instantons in quantum gravity are the analogues of tunnelling
solutions in quantum mechanics. When we consider tunnelling of a point
particle from an unstable minimum $x_{\infty}$ to a position $x_0$ of
zero momentum, we calculate the transition amplitude $<x_0\;, \;0\; |
x_{-\infty}, -\infty>$. One can show that the SOH is well approximated
by $Ae^{-S}$ where $A$ is a prefactor and $S$ is the action of the
classical Euclidean solution. By analogy, an instanton in gravity is a
solution of the Euclidean Einstein equations that interpolates between
an initial unstable state $U_0$ --approached asymptotically-- and a
zero-momentum hypersurface $U_1$ which is initial data for the post-decay
Lorentzian evolution. The existence of such an instanton is usually
taken as strong evidence that the transition takes place and the
amplitude is approximately given by $Ae^{-S}$ where $S$ is the action
of the instanton.  We are investigating the suggestion that the SOH in
quantum gravity be defined fundamentally as a sum over CCAL cobordisms
--or over AL cobordisms with causal continuity enforced
dynamically. That means first of all that there must be some CCAL
cobordisms for the transition under consideration. Secondly, it seems
reasonable that there would only be an instanton approximation if the
instanton had a background topology that was included in the sum over
manifolds in equation (1), {\it i.e.}, one which admits CCAL metrics.
Thus we want to check that when instantons are invoked as evidence
that topology changing processes occur, the instanton manifolds admit
CCAL metrics.
  
%             The instanton metric is invariant under reflection in the zero momentum
% hypersurface that initiates post-decay evolution. Hence the spacetime
% topology it describes is actually the double instanton or bounce: the
% cobordism that mediates the transition is any of the two the halves separated 
% by $U_1$. The boundary of the bounce should therefore contain two copies of 
% $U_0$ and a doubled side boundary\footnote{This sideboundary represents
% the union of the boundaries of all spacelike hypersurfaces in the time interval 
% between the initial and final ones.}, while $U_1$ only appears after the halving.

\subsection{Localised topology change}\label{localised}

Before turning to our specific examples, we first prove some results
necessary because the processes to be considered are embedded in an
ambient asymptotically flat region. We could think of this as the
topology change taking place within a lab with fixed walls
say. Clearly our Morse and handlebody technology will have to be
adapted to apply to these non-compact manifolds. This will not be
difficult because, with the assumption that the topology change is
localised in space, we can reduce the questions to the closed case by,
roughly speaking, closing off space. Once we demonstrate the existence of
CCAL metrics in the compact cobordism, we open back to
the physical manifolds. That this can be done without disrupting the Morse
structure of the metric is the content of the ``decompactifying" lemmas
stated below. Their proof is deferred to appendix \ref{metricappend}.
 
In the statement of lemmas \ref{lemmadec1} and \ref{lemmadec2}, 
we use the concept of a
gradient-like vector field for a Morse function $f$ on a manifold $M$.
Defining such a vector field amounts to covering $M$ with a congruence
of curves, along which $f$ increases, without reference to any
particular Riemannian metric on $M$. We borrow the definition from
Milnor (Lemma 3.2., \cite{milnor65}), while our construction of a concrete
vector field is a simple generalisation to non-elementary cobordisms
of the one given therein. Let $f$ be a Morse function on the
$n$-dimensional manifold $M$ with a set of $r$ Morse points
$P=\{ p_k\} $. For simplicity we assume that each Morse point occurs
on a distinct level surface of $f$ though this assumption can easily
be dropped.

        A {\it gradient-like vector field} $\xi $ for $f$ is a
        smooth vector field on $M$ with properties:

        (i) $\xi (f) > 0 \;\; \forall q \notin P$

        (ii) $\xi$ has coordinates $(-2x_1, \cdots, -2x_{\lambda _k},
        2x_{\lambda _k +1}, 
\cdots, 2x_n)$ in a neighbourhood of $p_k$ where $f$
        admits expansion \scalebox{.85}[1]{$f(q) = f(p_k) -
        \sum\limits_{1\leq i\leq \lambda _k}x_i^2 +
        \sum\limits_{\lambda_k< j\leq n}x_j^2 $}     
 
        A vector field satisfying these two conditions can always be
        found in \nolinebreak $M$.  Indeed, pick an atlas $\mathcal{A} =
        {(U_{\alpha}, \phi_{\alpha})} \; \alpha =1, \cdots N$ so that
        a single chart $U_k$ contains the critical point $p_k$ and so
        that, dividing the range $\{ \alpha \} $ as $\{ k, a\} \;\;
        k=1,\cdots , r \;\; a= r+1,\cdots, N$, the following
        hold:
\begin{enumerate} 
        \item For each k, there is a smaller neighbourhood $U'_k
\subset U_k$ satisfying
\begin{eqnarray} \bar{U}'_k \cap U_{\alpha} &=& \left\{ \begin{array}{ll}
                                 \bar{U}'_k & \mbox{if $k=\alpha$} \\
                                 \emptyset & \mbox{otherwise}   \end{array} \right.
\nonumber \\
\mbox{and} \;\;\;\phi_k(U'_k) &=& \{\vec{x}\in \mathR ^n : |\vec{x}|^2<
\varepsilon\}\;\; \mbox{for some small} \;\;\varepsilon \nonumber
\end{eqnarray} 
where $\bar U_k'$ means the closure of $U_k'$.
\item In $U_k$ $f$ has local representative
\scalebox{.85}[1]{$f_k(\vec{x})\equiv f\circ\phi ^{-1}_k(\vec{x})=c_k -\sum
\limits_{1}^{\lambda _k}x_i^2 +\sum\limits_{\lambda_k+1}^{n}x_i^2$}
        \item In $U_a$ $f$ has local representative $f_a(\vec{x})\equiv f\circ\phi ^{-1}_a(\vec{x})=\mbox{const}
+x_1^{(a)}$

\end{enumerate}

        We now define $\xi$ chart by chart. In $U_k$ we give it the
        components \linebreak $\xi ^{(k)} = (-2x_1, \cdots, -2x_{\lambda
        _k}, 2x_{\lambda _k +1}, \cdots, 2x_n)$ and in $U_a$
        $\xi ^{(a)} = (1, 0, \cdots, 0)$. Then we combine
        the local representatives $\xi ^{(\alpha)}$, through a partition
        of unity $\{\theta _{\alpha}\}$ for 
        $\mathcal{A}$ to obtain a 
        vector field, $\xi = \sum_\alpha \theta_\alpha \xi^{(\alpha)}$,
which clearly satisfies condition (i) and condition (ii)
in the neighbourhood $U_k'$ of $p_k$.

Covering the case of ordinary asymptotic flatness we have
        \begin{lemma}\label{lemmadec1}
        Consider two non-compact asymptotically flat (n-1)-geometries 
$(U_0, h_0)$ and $(U_1, h_1)$.  Suppose that the closed manifolds 
$V_0$ and $V_1$ are one-point compactifications of $U_0$ and $U_1$, in the 
sense that there are points $\tilde q_i \in V_i$ and 
diffeomorphisms $\tilde{d}_i: V_i -\tilde q_i\rightarrow U_i$, $i = 0,1$.
Further suppose that there is a triad $(M,V_0,V_1)$ with a Morse
function $f:M\rightarrow [0,1]$ with no index $n$ critical points.

        Then,

(i) There is an integral curve $\mathbf{C}$ of a gradient-like vector
field for $f$ which traverses $M$, from $V_0$ to $V_1$ without
intercepting any critical point.

(ii) The manifold $L\equiv M-\mathbf{C}$ is a cobordism between $U_0$
and $U_1$ and there is an AL metric on $L$ which has the same Morse
structure as $f$, is asymptotically flat and has the 
correct restrictions,
the pull-backs of $h_0$ and $h_1$, on the boundary $\partial L =
(V_0 - q_0) \uplus (V_1 -q_1)$ where $ q_i = V_i \cap\mathbf{C}$.
\end{lemma}

For asymptotic Kaluza-Klein boundary conditions consider
compactifying $\mathR ^3\times S^1$, the topology of a spatial section in
the 5-dimensional Kaluza-Klein vacuum: we add a whole circle, one
point at infinity of $\mathR ^3$ for each point of $S^1$. In the reverse
process an $S^1$ must be removed to recover the physical boundaries
from the closed manifold. While all points in a
manifold are equivalent, in general not all embedded circles are:
given a manifold $V$, the manifolds $V-C$ and $V-\tilde{C}$ may not be
diffeomorphic if $C$ and $\tilde{C}$ are different embedded circles. 
In order to decompactify to Kaluza-Klein boundary
conditions along the lines of \ref{lemmadec1}, we enlarge our list of
hypotheses with a further condition which guarantees the equivalence
of all subtracted circles in the closed boundary \nolinebreak $V_1$.
        \begin{lemma}\label{lemmadec2} 
Consider two asymptotically Kaluza-Klein flat
        (n-1)-geometries $(U_0,h_0)$ and $(U_1,h_1)$. Suppose that there exist 
closed manifolds $V_0$ and $V_1$, with $V_1$ connected and simply connected, 
and diffeomorphisms $\tilde{d}_i: V_i - \tilde{C}_i\rightarrow\nolinebreak U_i$
 with $\tilde{C_i}\subset V_i$ diffeomorphic to $S^1$, $i = 0,1$.  
Further suppose that there is a triad 
$(M,V_0,V_1)$ with a Morse function $f:M\rightarrow [0,1]$ with no index 
$n$ or $(n-1)$ critical points.

        Then,

(i) There is an ``integral annulus" $\mathbf{A}$ for the gradient-like
vector field $\xi$ --- by this we mean an $S^1$ worth of integral curves
of $\xi$, {\it i.e.}, an imbedding $i: B^1\times S^1 \hookrightarrow M$
such that for each point in the circle $\psi \in [0,2\pi )$ the
segment $i(B^1\times \{\psi\}) $ is an integral curve of $\xi $ --- which
traverses $M$, from $V_0$ to $V_1$ without intercepting any critical
point.

(ii) The manifold $L\equiv M-\mathbf{A}$ is a cobordism between $U_0$
and $U_1$ and there is an AL metric on $L$ which has the same Morse
structure as $f$, is asymptotically flat and has correct restrictions,
the pull-backs of $h_0$ and $h_1$, on the boundary $\partial L = (V_0 - C_0)
\uplus (V_1 -C_1)$ where $C_i=V_i\cap\mathbf{A}$, i=0,1.
\end{lemma}

        \subsection{Pair production of black holes} \label{bhpair}
Due to the positive energy theorems, the Minkowski vacuum $M^4$ is stable 
with respect to semi-classical decay \cite{witten82}. However a cylindrically 
symmetric magnetic field described by the Melvin solution can decay into a pair
of oppositely charged black holes thanks to the extra energy contained in the 
field \cite{strominger91,gibbons86}. The instanton that governs the decay is the Euclideanised 
Ernst solution.  To see the topologies associated with the metrics involved, the 
reader is encouraged to consult \cite{strominger91}. We take them as the 
starting point for analysing the cobordism. They are a spacelike hypersurface 
of Melvin, $\mathR ^3$, a post-tunnelling spacelike hypersurface containing a pair of 
black holes, $S^2 \times S^1 -\{point\} $ and the doubled instanton, 
or ``bounce'' topology, $S^2 \times S^2 -\{point\}$. Removing a point from 
a 4-dimensional closed manifold is equivalent to removing a closed ball $B^4$:
it gives a non-compact manifold. We compactify by adding the  point back 
in and cut the bounce in half to obtain $\underline{M} \cong S^2 \times B^2$. We 
then delete an open four-ball to create the initial boundary. The manifold 
$\underline{M}- \dot{B}^4$ is $M$ in Lemma 1, $V_0 \cong S^3$ is the initial 
boundary and $V_1 \cong S^2\times S^1$ is the final boundary.

\begin{figure}[ht]
\centering
\rotatebox{270}{\resizebox{!}{3.5in}{\includegraphics{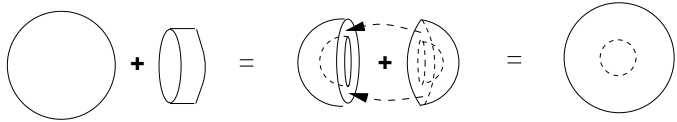}}}
\caption{\protect\parbox[t]{4.5cm}{\small Handlebody for $S^2\times B^1$: a ball with a hole at the
centre.}\label{hols}}
\end{figure}

        Combining Fig \ref{hols} with right-distributivity gives the following 
handlebody decomposition for $\underline M$ :
\begin{eqnarray*}
\underline{M} &=& S^2\times B^2 \cong (S^2\times B^1)\times B^1 \cong
(B^3 + B^2\times B^1)\times B^1 \\ &=& B^4 + B^2\times B^2
\end{eqnarray*}

Thus there exists a Morse function on $M$ which contains only a Morse
point of index 2. Dowker and Surya gave an earlier proof by explicitly constructing an allowed 
Morse function \cite{surya} on $\underline{M}$ that can, in fact, be regarded as
associated with the handlebody decomposition given above. 
Lemma  \ref{lemmadec1} then shows that there is an asymptotically flat 
CCAL metric on the non-compact cobordism $L=M-\mathbf{C}\cong 
S^2\times B^2 - \dot{B}^4 - B^1 $ where $\mathbf{C}$ is an integral 
curve of a gradient-like vector field of the Morse function. $L$ is diffeomorphic 
to the original cobordism --half of ($S^2 \times S^2 - \{point\})$-- up to the observation
that the initial boundary in $L$ is at some finite time in the past whereas 
in the original cobordism it is in the infinite past.

\begin{figure}[ht]
\centering
\resizebox{!}{2.5in}{\includegraphics{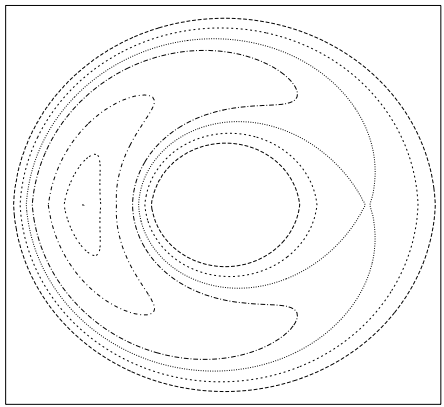}}
\caption{\protect\parbox[t]{8.3cm}{\small Levels of a Morse function $f$ in the cobordism $S^2 \rightarrow S^1 \times S^1$. On the left, an index 0 point accounts 
for the creation of $S^2$; on the right an index 1 point marks the transition to $S^1\times S^1$. The function $f$ increases in the direction of the expanding  spheres and then in the direction of the expanding tori.  
Critical character of the Morse points is reflected in a same behaviour of $f$
along a Cartesian direction and its opposite.}\label{bhlevels}}
\end{figure}

        We can illustrate the location of the critical points and the critical levels in 
a lower dimension, $n=3$. The equivalent process would be 
$S^2 \rightarrow S^1 \times S^1$, which is mediated by part of the handlebody 
$B^2 \times S^1 = B^3 + B^1 \times B^2$, and hence contains an unwanted 
index 1 point. The two critical points lie in the interior of the solid torus
$B^2\times S^1$ as depicted in Fig \ref{bhlevels}, a two dimensional
section.

        This construction generalises to higher dimensions, so that black hole 
pair creation is feasible whenever $n \geq 4$ as shown in \cite{surya}. 
Indeed, applying right distributivity 
of the $B^2$ ball to $S^{n-2} = B^{n-2} + B^{n-2}$ we obtain:
\begin{equation}
S^{n-2} \times B^2 = B^n + B^{n-2} \times B^2
\end{equation}
Thus the cobordism $S^{n-1} \rightarrow S^{n-2} \times S^1$ contains
only an index $(n-2)$ critical point, which respects causal continuity
whenever $n\geq 4$.

\subsection{Decay of Kaluza-Klein vacuum and magnetic
field}\label{kkmonopole}
        
        In five-dimensional Kaluza-Klein gravity a fifth compact dimension is 
added to ordinary 4-dimensional spacetime. The corresponding metric has 
fifteen degrees of freedom, which can be interpreted as one dilaton scalar, four 
components of the electromagnetic field and ten components of the spacetime
 4-metric. The 4-dimensional spacetime associated 
with a given 5-geometry is obtained by reduction along a Killing vector field of 
closed orbits. Both the Kaluza-Klein vacuum and the Kaluza-Klein
version of the Melvin solution have a 
background topology $\mathR ^4\times S^1$ and are semi-classically unstable 
\cite{witten82,gauntlett94,gauntlett95}.

        Now the relevant instanton is the euclideanisation of a five-dimensional 
black hole: 5-d Schwarzschild for the decay of the vacuum and a rotating 
5-d Kerr solution for the decay of a magnetic field. The latter is interpreted, 
upon four dimensional reduction, as pair production of Kaluza-Klein monopoles. 

        The topology change is the same in both cases, from the unstable 
spacelike hypersurface $ \mathR ^3\times S^1$
to the starting hypersurface for post-decay $ \mathR ^2\times S^2\cong S^4-S^1$. 
The double instanton has topology $\mathR ^2\times S^3 \cong S^5 - S^1$.
Once more we compactify by replacing
 the circle and then halve the closed $S^5$ bounce to get 
$\underline{M}\cong B^5$, with $\partial B^5 = S^4$. Finally we delete an open
thickened circle $S^1\times \dot{B}^4$ from $\underline{M}$ to create the initial
boundary. This yields $M$ with $\partial M = S^1\times S^3
\uplus S^4$.  That is, with the notation of Lemma 2, we have the triple 
$(M, V_0, V_1) = (B^5-S^1\times \dot{B}^4, S^3\times S^1, S^4)$. 
We seek a handlebody decomposition for $B^5$ which truncates into a 
cobordism from $S^3\times S^1$ to $S^4$. 
The ``redundant'' $B^3$ decomposition eq.(\ref{longdec}) and 
right-distributivity imply the identity:
\begin{eqnarray*}
B^5 &=& B^3\times B^2 \\ &=& (\underbrace{B^3 + B^1\times
B^2}_{B^2\times S^1} + B^2\times B^1)\times B^2 =\underbrace{B^5 +
B^1\times B^4}_{B^4\times S^1} + B^2\times B^3
\end{eqnarray*}

The first term, $B^5$, corresponds to the creation of $S^4$ from 
$\emptyset$ and 
the first handle addition corresponds to the transition from $S^4$ to
$\partial(B^4 \times S^1) = S^3 \times S^1$, {\it i.e.}, the (closed) 
KK vacuum space. The second handle addition is therefore the one that
corresponds to the process we are investigating, $S^3 \times S^1
\rightarrow
S^4$. This means that in the cobordism between $S^1\times S^3$ and $S^4$,
which involves only the handle $B^2\times B^3$, there is a Morse function with
exactly one critical point of index 2, i.e., no index 1 or 4 points.

We apply Lemma \ref{lemmadec2} to $M$ and conclude that since
the original instanton manifold (half of $S^5-S^1$) 
is diffeomorphic to  $L = M -
{\mathbf{C}}$ 
 there exist asymptotically CCAL metrics on it.

The next figure represents a section of the 3-dimensional
analogue of the cobordism $M$. The whole cobordism between $S^1\times S^1$
and $S^2$ is generated by revolution around the z-axis. The reader can try and
picture a cylinder between the inner boundary and the outer boundary that is 
orthogonal to the contours and does not touch the critical point at the centre.
That would be the integral annulus $\mathbf{A}$ of Lemma \ref{lemmadec2}.

\begin{figure}[ht]
\centering
\resizebox{!}{2.5in}{\includegraphics{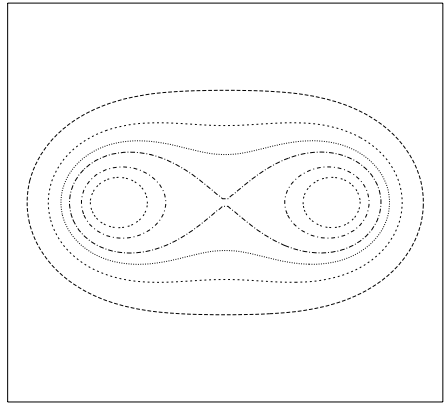}}
\caption{\protect\parbox[t]{8.3cm}{\small Levels of the Morse function that represents the change $S^1 \times
S^1\rightarrow S^2$. Notice that this is in fact Figure \ref{bhlevels} turned inside out: the time reversed cobordism if time progresses from inner to outer surfaces . Thus the central point, which
was there an index 1 point is here an index 2 point: the Morse function increases
in the z-direction and decreases in the other two.}\label{kklevels}}
\end{figure}

This result also generalises to a countable family of
higher-dimensional cobordisms that mediate the nucleation of various p-branes
\cite{gauntlett96}. In n-dimensional Kaluza-Klein theory the Kerr instanton
manifold $\mathR ^2\times S^ {n-2} \cong S^n - S^1$ is the double of
the cobordism that mediates the transition $\mathR ^{n-2}\times S^1 \rightarrow
\mathR ^2\times S^ {n-3}$,
which in the closed case reads $S^{n-2}\times S^1 \rightarrow S^
{n-1}$.These are respectively the boundaries of $B^{n-1} \times S^1$
and $B^n$; since
\begin{eqnarray}
B^n &=& B^3 \times B^{n-3} = \nonumber \\ 
&=& B^n + B^1\times B^{n-1} + B^2 \times B^{n-2} 
\end{eqnarray}
Again, it is the second handle addition that corresponds to the
process of interest and we see that there exists a Morse function with
only one critical point of index 2, which respects causal continuity 
when $n\ge 4$.

\section{Discussion}\label{discussion}

We have seen that several instantons for interesting topology changing
processes have topologies that 
support Morse metrics without index 1 or $n-1$ points. 
These include the pair production of black holes in any dimension,
the decay of the Kaluza-Klein vacuum and pair production of
Kaluza-Klein monopoles. We have called such cobordisms CCAL based on the
conjecture that causal discontinuity of Morse metrics
is associated with and only with indices 1 and $n-1$. This conjecture
remains to be proved.

Further work that would place our results on a more physical footing 
would include the study of quantum fields propagating on Morse 
metric backgrounds with different indices. Can a quantum field 
be non-singular on a spacetime with an index 2 point in 4 spacetime
dimensions for example?

Another question is  whether the 
equivalence relation of ``cobordant via a CCAL cobordism'' results in
a finer subdivision of $(n-1)$-manifolds than simple cobordism.  
What we know is that for $n-1 = 3$ this doesn't happen. Any two
3-manifolds are cobordant and also cobordant via a CCAL cobordism
\cite{surya}.
We do not know what
the  situation is in higher dimensions. 

The higher dimensional question is interesting for the 
following possibility. 
As far as simple cobordism goes any topology of the form 
$S^{k} \times A$ where $A$ is any closed manifold is possible for
space since it is cobordant to
the empty set via the cobordism $B^{k+1}\times A$.
If it is right to restrict to CCAL cobordisms in the SOH then 
        this might constrain the possible topologies of the universe
because some
may not be CCAL cobordant to $\emptyset$. $A$ might be some Calabi-Yau
manifold for example or a torus $T^{n-k}$ so CCAL cobordism might 
restrict some of the possible string theory  compactifications if the 
universe was created from nothing as is usually conceived.

We end on a cautionary note. 
The idea that only CCAL cobordisms be included in the SOH for quantum
gravity must be scrutinised in the light of the result that the 
canonical ``U-tube'' cobordism for pair production of topological 
geons is not CCAL \cite{surya}. This would leave in serious trouble the 
proposal that a spin statistics correlation for quantum geons can 
be established by a ``topological'' argument similar to that for 
skyrmions \cite{sorkin89}\cite{fay96}.  
On the other hand if it is true that there's an underlying discrete
substructure to spacetime, it will likely regulate infinities such
as the singular quantum field behaviour. In that case 
the suppression of non-CCAL cobordisms will be only finite and 
they would need to be included after all.

\bigskip
{\bf Acknowledgements}
\medskip
We would like to thank A. Chamblin, G. Gibbons, C. Isham, R.Penrose,
R. Sorkin, S. Surya, P. Tod and N. Woodhouse
for help and interesting discussions. H.F. Dowker is supported in 
part by an EPSRC Advanced Fellowship.

\newpage

\appendix
\section{Combining manifolds with boundary.}\label{m&gmanappend}

        We indicate the difficulties in defining a differentiable
        structure in the product or adjunction of two manifolds with
        boundary  --both operations are essential in the construction of
        handlebodies. What makes this problematic is the requirement
        that coordinate charts in a $\partial $-manifold\footnote{Throughout 
this appendix $\partial$- manifold means compact manifold with boundary.} $M$ be
        homeomorphisms from a neighbourhood of a point $p \in M$ onto
        an open set of the ``hemiplane'' $H^n \equiv \{\vec{x} \in \mathR ^n
        : x^n \ge 0\}$. The irrelevance of the order in which atlases
        are defined makes the map $f$ in eq.(\ref{diffeo}) a
        diffeomorphism.

        {\bfseries Product of two $\partial$-manifolds.} An atlas in
        the product of two closed manifolds --or in the product of
        a closed manifold and a $\partial $-manifold-- $M$
        and $N$, of respective dimensions $m$ and $n$, is naturally
        induced by the individual coordinate charts of the two
        manifolds. In the usual notation, given a chart
        $(U_{\alpha},\phi_{\alpha})$ around the point $p \in M$ and a
        chart $(V_{\beta},\psi_{\beta})$ around $q \in N$, the chart
        $(U_{\alpha}\times V_{\beta},\phi_{\alpha}\times
        \psi_{\beta})$ maps a neighbourhood of $(p,q) \in M\times N$
        homeomorphically onto an open subset of $R^{m+n}$ and
        differentiability of the transition functions between charts
        automatically follows from that in $M$ and $N$. However, for
        the product of two $\partial $-manifolds the set
        $\phi_{\alpha}\times \psi_{\beta}\; (U_{\alpha}\times
        V_{\beta})$ is not open in $H^{m+n}$ when $U_{\alpha}\times
        V_{\beta}$ contains points in $\partial M\times \partial N
        \subset \partial (M\times N)$. The
        troublesome region $\partial M\times \partial N$ is called the
        {\it corner set} of $M\times N$. Standard $\partial
        $-manifold charts can also be defined in this region by fixing
        a homeomorphism $H^1\times H^1 \rightarrow H^2$, see
        \cite{brocker73}. Then using the collaring theorem the
        $(\partial M\times H^1)\times (\partial N\times H^1)$ is
        identified with $\partial M \times \partial N\times H^2$,
        which is embedded in $M\times N$ as a usual collar and
        thus a genuine part of the boundary. It is this ``deformed''
        structure, where an atlas can be properly defined, what we
        mean by the product of two $\partial $-manifolds.

        {\bfseries Attachment of $\partial $-manifolds through a
        boundary identification.} Down in the category of topological
        spaces, handle attachment is a special case of adjunction of
        two spaces. The adjunction of spaces $X$ and $Y$ through a map
        $h: A\subset X \rightarrow Y$ is the quotient space of the
        disjoint union $X\uplus Y$ by the equivalence relation
        $R_h$. Given two points $z, z' \in X\uplus Y$,
\[
 z\; \cong _{R_h} z'\; \; \; \mbox{iff}\; \; \; \left\{
 \begin{array}{rl} z&=\; z'\\ z &=\; h(z')\\ z' &=\; h(z)\\ h(z) &=\;
 h(z') \end{array}\right.
\]

The adjunction space, denoted $ \frac{X \uplus Y}{R_h}$ or simply $X
+_h Y$, is endowed with a topology \cite{maunder70} through the
projection $\pi : X\uplus Y \rightarrow X +_h Y$. Points in $X +_h Y$
are the equivalence classes $[z]$ of $R_h$.

        We attach a handle $H$ to a manifold $M$ through an injection
        $h$ from a closed subset $A \subset \partial H$ onto a region
        of $\partial M$ diffeomorphic\footnote{In this case the fourth
        possibility for equivalence can be omitted; it implies the
        first by $h$ injectivity.} to $A$ . A $\partial $-manifold
        structure is induced in $H +_h M$ with an atlas $\mathcal{A}$
        such that $\partial A \cong \partial (h(A)) $ remains in the
        boundary but $\dot{A}$ becomes part of the interior of $H +_h
        M$. On defining the charts in $\mathcal{A}$ from those in the
        atlases of $H$ and $M$, the region $\partial A$ behaves as a
        corner set and smoothing is required again. Such smoothing is
        implicit when we talk of the manifold $M +_h H$. A 
        justification of the glueing process, that uses the collaring
        theorem can be found in \cite{kirby77,brocker73}.

        The map $f: (L +_h B^\lambda\times B^{n-\lambda})\times B^1\rightarrow
        L\times B^1 +_{\tilde{h}} B^\lambda\times B^{n+1-\lambda} $ of
        eq.(\ref{diffeo}) essentially reverses the order in which
        product and adjunction are performed. Continuity of $f$ and
        its inverse is easily checked with respect to the topologies
        on either side. Finally, to establish differentiability of $f$
        as a map between the two smoothed manifolds we would have to
        display the modified local charts and combine $f$ with the
        smoothing operations, an exercise in differential topology
        that would take us too far afield.
\section{ A.L. Metrics with correct boundary
conditions.}\label{metricappend}

{\bf Proof of lemma \ref{lemmametric}.} 
Theorem \ref{morsexistT} tells us that there is a Morse function
$f:M\rightarrow [0,1]$ with:
\begin{enumerate}
        \item $V_0 = f^{-1}(0)$ and $V_1 = f^{-1}(1)$ \item $f$ has no critical
points on $\partial M = V_0 \uplus V_1$
\end{enumerate}
 
        We construct a Morse metric $g$, associated with
        $f$, such that $g_{|_{V_i}} = \tilde h_i \equiv 
        d^*_i(h_i) \quad i=0,1$.
Our proof is divided in two steps:
        first we find conditions on a Riemannian metric $G$ 
        that are sufficient for $g$ defined by  eq.(\ref{eq.morse.g}) to 
        have the correct restrictions; then we show that a Riemannian
        metric exists on $M$ satisfying those conditions.

 When a spacelike
        hypersurface $V \subset M$ is defined as a level surface of a function
        $f:M \rightarrow\mathR$, the restriction of a metric $g$ on $M$ to $V$ is given by \nolinebreak: \be
\label{induced}
h^{(g)}_{\mu \nu} = g_{\mu \nu } \mp
\frac{\partial_{\mu}f\partial_{\nu}f}{|g^{\rho \sigma}
\partial_{\rho}f\partial_{\sigma}f|} \ee

\noindent where the $-$ or $+$ sign
applies respectively to the cases of Riemannian or Lorentzian $g$. So if we
wish to compute the restriction of the Morse metric $g_{\alpha \beta
}$ to the boundary manifolds $V_0 \equiv \{x \in M : f(x)=0\}$ and
$V_1 \equiv \{x \in M : f(x)=1\}$ we need its inverse $g^{\alpha
\beta}$. From
\begin{displaymath}
g_{\mu \nu} = \partial _{\rho} f\partial _{\sigma}f\; G^{\rho
\sigma}G_{\mu \nu} - \zeta \partial _{\mu } f\partial _{\nu }f
\end{displaymath}
\noindent one easily finds that:
\[
g^{\mu \nu } = \frac{1}{(\partial f)^2}\; G^{\mu \nu } + \frac{\zeta
}{(\partial f)^4(1-\zeta)}\; G^{\mu \rho}G^{\nu \sigma }\partial
_{\rho } f\partial _{\sigma }f
\]
 
\noindent where $(\partial f)^2 $ denotes $G^{\rho \sigma} \partial_{\rho}
f\partial_{\sigma}f$. The $g$-induced metric
on $V_i$, $i = 0,1$, is \be
\label{g-induced}
h^{(g)}_{i\mu \nu} = g_{\mu \nu } +
\frac{\partial_{\mu}f\partial_{\nu}f}{|g^{\rho \sigma}
\partial_{\rho}f\partial_{\sigma}f|} \ee
So evaluate
\begin{eqnarray*}
g^{\rho \sigma} \partial_{\rho}f\partial_{\sigma}f
&=&\left(\frac{1}{(\partial f)^2}G^{\rho \sigma} + \frac{\zeta
}{(\partial f)^4(1-\zeta)}G^{\rho \alpha}G^{\sigma \beta}\partial
_{\alpha} f\partial _{\beta}f\right)\partial_{\rho}f\partial_{\sigma}f
\\ &=& \left(1 + \frac{\zeta }{(\partial f)^4(1-\zeta)}(\partial
f)^4\right) = 1 + \frac{\zeta}{1-\zeta} = \frac{1}{1-\zeta }
\end{eqnarray*}
and insert this in eq.(\ref{g-induced}) to obtain
\bea h^{(g)}_{i\mu \nu} &=& (\partial f)^2G_{\mu \nu } - \zeta
\partial_{\mu} f\partial_{\nu} f + |1 -\zeta | \partial_{\mu }
f\partial_{\nu} f\nonumber \\ &=& (\partial f)^2G_{\mu \nu } -
\partial_{\mu } f\partial_{\nu } f = (\partial f)^2 \left(G_{\mu \nu}
- \frac{\partial_{\mu } f \partial_{\nu }f}{(\partial f)^2} \right)
\nonumber \\ &=&(\partial f)^2h^{(G)}_{i\mu \nu} \nonumber \eea

From this equation we immediately infer sufficient conditions on $G$.
If G restricts to $h^{(G)}_i = \tilde h_i$ and is 
such that $(\partial f)^2 = 1$ on $V_i$, then:
\begin{displaymath}
 h^{(g)}_i= h^{(G)}_i = \tilde h_i
\end{displaymath}
and we will be done.

These two conditions can be fulfilled by adapting $G$ to the
level surfaces of $f$, in the sense of Theorem 3.1. in
\cite{milnor63}. More explicitly, by compactness and isolation of the
Morse points $\{ p_k\} $, $k = 1,\dots r$,
the collaring theorem \cite{stong68,milnor65} permits the following
factorisation of an open neighbourhood $M_0\subset M$ of $V_0$:
\[
        M_0 \equiv \{p\in M :0\le f(p) <\epsilon_0\}
\stackrel{\Phi _0}{\cong} V_0 \times [0 , \epsilon_0) 
\]
where $\epsilon_0 < c_1$ and $f(\Phi _0^{-1}({x}, t)) = t $ for all 
$({x},t)$ in $V_0 \times [0,\epsilon_0)$. This foliation
of $M_0$ is adapted to $f$ in the sense that $\partial_t f = 1$ and
$\partial_{x^i} f = 0$ where $x^i$, $i= 1,\dots n-1$, are some local 
coordinates on $V_0$ and thence on $M_0$.  
Similarly let $M_1 \equiv \{p\in M: 1-\epsilon_1 < f(p) \le 1\} 
\stackrel{\Phi _1}{\cong} V_1 \times (1-\epsilon_1 , 1]$ where $c_r < 1-
\epsilon_1$. Let $M_2 = f^{-1}((\delta, 1-\delta))$ where $\delta = 
{1\over 2} {\rm min}(\epsilon_0, \epsilon_1)$.  
 We can then express M as the union of open subsets: 
$M =M_0 \cup M_2  \cup M_1$. Next we define a Riemannian metric 
$G_0$ on $M_0$ as the $\Phi _0$ pull-back of the metric on 
$V_0 \times [0 , \epsilon_0)$ with interval 
$ds^2 = dt^2 + \tilde h_0(x)_{ij}\;dx^idx^j$ 
and similarly we define $G_1$ on $M_1 \cong V_1 \times (1-\epsilon_1,
1]$. Take an arbitrary Riemannian metric $G_2$ on $M_2$. One 
certainly exists by paracompactness of $M_2$ as a submanifold of the 
compact manifold $M$.

Now let $(U_{\alpha}, \phi_{\alpha})$ be a finite atlas for $M$, $\alpha= 1,\cdots , N$
say, then the sets $W_{\alpha\, 0}\equiv U_{\alpha} \cap M_0$, $W_{\alpha 1}\equiv
U_{\alpha}
\cap M_1$ and $W_{\alpha 2}\equiv U_{\alpha} \cap M_2$ give another finite
cover of $M$: a refinement of the atlas 
$(U_{\alpha}, \phi_{\alpha})$. Using an associated partition of unity 
$ \{ \theta _{\alpha i}\}$ we can construct a metric 
on $M$ by patching together local metrics \cite{dubrovin95}; so define 
\[
        G_{\mu \nu} (p) = \sum_{\alpha i} \theta _{\alpha i} (p)\;G_{\mu \nu}^{\alpha i}(p)      
\]
\noindent with $\hspace{3.95cm} G_{\mu \nu}^{\alpha i}  = (G_{i \,|W_{\alpha i}})_{\mu \nu} $ 

\noindent Throughout $M_0 - M_2\cap M_0$, a neighbourhood of $V_0$, we have:
\[ G^{\mu \nu} \partial_{\mu} f\partial_{\nu}f =
\underbrace{\partial_0 f}_{ = 1}\partial_0 f + \tilde h_0^{ij}
\underbrace{\partial_i f}_{= 0}\partial_j f = 1 \nonumber \] 
and similarly in a neighbourhood of $V_1$. In particular, the
unit-norm condition is satisfied on $V _0  \uplus  V_1$ and
substituting in the expression for the induced metric $h^{(G)}_{0\mu
\nu} = G_{\mu \nu } - \frac{\partial_{\mu}f\partial_{\nu}f} {G^{\rho
\sigma}\partial_{\rho}f\partial_{\sigma}f}$ gives:
\begin{eqnarray}
 h^{(G)}_{0\, 00} &=& G_{00}-\partial_0 f\partial_0 f = 0 \nonumber \\
 h^{(G)}_{0\, 0i} &=& G_{0i}-\partial_0 f\partial_i f = 0 \nonumber \\
 h^{(G)}_{0\, ij} &=& G_{ij}-\partial_i f\partial_j f =
\tilde h_{0\, ij}\nonumber
\end{eqnarray}

\noindent so that $G_{|_{V_0}} =\tilde{h}_0$. 
Similarly $G_{|_{V_1}} = \tilde{h}_1$ . Hence the result.
 $\Box $

       {\bf Proof of Lemma \ref{lemmadec1}.}
 We will find a solid cylinder $C$ of such integral
        curves and take $\mathbf{C}$ to be the central curve. It is
        convenient to start by defining the past and future ``shadows"
        of each Morse point $p_k$ in M.  To do so we work with the
        covering of $M$ used in the construction of the vector field
        $\xi $. Relabel the coordinates in the critical chart $U_k$ by
        setting $y_i=x_i \quad 1 \leq i \leq \lambda _k$ and $z_j =
        x_{\lambda _k +j} \quad 1\leq j \leq n-\lambda _k$. One can
        then verify \cite{milnor65, wallace68} that the integral curve of 
    $\xi$ through $(\vec{y},\vec{z}) \in \phi _k(U'_k)$ is
\begin{enumerate}
      \item A straight line ending at the origin,
          $\phi_k(p_k)=\vec{0}$, if $\vec{z} =\vec{0}$.

        \item A straight line beginning at the origin if $\vec{y}=\vec{0}$.

        \item A hyperbola which does not pass through the origin if
neither $\vec{y}$ nor $\vec{z}$ vanish; this hyperbola intersects the
$S^{n-1}$ boundary of $\phi _k(U'_k)$ in two points, one in $
-|\vec{y}|^2 +|\vec{z}|^2 = -\varepsilon $ and another in
$-|\vec{y}|^2 + |\vec{z}|^2 = \varepsilon $.

\end{enumerate}
        The past and future shadows of $p_k$ are defined to
        be the regions:
\[ \begin{array}{l}
        S^{(k)}_P \equiv \{q \in M : \exists \;s>0 \mbox{~~ with ~~}
        \sigma _q^{\xi }(s)=p_k\} \\
        \\ S^{(k)}_F \equiv \{q  \in M: \exists \;s<0 \mbox{~~ with ~~}
        \sigma _q^{\xi }(s)=p_k\} 
        
\end{array} \]
where $\sigma^{\xi}_q$ is the integral curve of $\xi$ which passes
through q at $s=0$, {\it i.e.}, $\sigma^{\xi}_q (0)=q$ and the
parameter $s$ is fixed by $\xi (f)\, (q)=
\frac{df\circ\sigma^{\xi}_q}{ds}_{| s=0}$. Then choose for each
k a pair of real numbers $a_k,\; b_k$ close enough to the critical
value, $0 < c_k -a_k < \varepsilon$ and $0 < b_k - c_k < \varepsilon$, so that
the intersection of the past and future ``shadows" of $p_k$ with
respectively $V_{a_k}\equiv f^{-1}(a_k)$ and $V_{b_k}\equiv f^{-1}(b_k)$ are 
complete spheres. Specifically:

\[\begin{array}{l}
        S^{(k)}_L \equiv V_{a_k}\cap S^{(k)}_P = \phi _k^{-1}\left( \{
        (\vec{y}, \vec{z}) : \vec{z} = \vec{0} \; \mbox{and}\;
        |\vec{y}|^2=c_k - a_k\} \right) \cong S^{\lambda _k -1}\\ \\
        S^{(k)}_R \equiv V_{b_k}\cap S^{(k)}_F = \phi _k^{-1}\left( \{
        (\vec{y}, \vec{z}) : \vec{y} = \vec{0} \; \mbox{and} \;
        |\vec{z}|^2=b_k - c_k\} \right) \cong S^{n -\lambda _k -1}
        
\end{array} \]
        The left sphere $S^{(k)}_L$ includes all the points of
        $V_{a_k}$ in curves that end in $p_k$, while the right sphere
        $S^{(k)}_R$ includes all the points of $V_{b_k}$ which begin
        at $p_k$. By considering just the left spheres, we can now show 
        that if the index of
        each critical point $p_k$ satisfies $0 \leq \lambda _k
        < n$, then there is an embedded solid cylinder $C$ in $M$
        diffeomorphic to $B^{(n-1)} \times B^1$ which contains no
        critical points and such that the $B^1$ coordinate of a point
        $p\in C$ is the value $f(p)$.

        We proceed by induction. It is convenient to define $a_{r+1}=1$ 
        so that $V_{r+1}\equiv V_1$.
        Suppose that we have found an integral cylinder $C_k$ from
        $V_0$ to $V_{a_k}$ with no critical points. It intersects the
        (n-1)-dimensional manifold $V_{a_k}$ in a disk $D_k \cong
        B^{n-1}$. We extend this cylinder forward to an integral
        cylinder from $V_0$ to $V_{a_{k+1}}$.

        Induction starts, because $V_{a_1} \cong V_0 $ and we can
        choose any disk $D_0 \in V_0$ and project it forwards along
        the $\xi$ -lines until another $D_1\cong D_0$ is reached in
        $V_{a_1}$. The whole set of $\xi$-lines from $D_0$ to $D_1$ is
        the integral cylinder $C_1$.

\begin{figure}[ht]
\centering
\rotatebox{270}{\resizebox{!}{2.5in}{\includegraphics{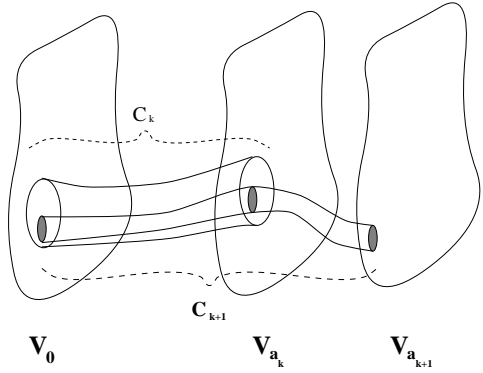}}}
\caption{{\small Extension of the cylinder.}\label{cyl}}
\end{figure}

To demonstrate how a typical inductive step works, we use the fact
that the $\xi$-curves ending at $p_k$ intersect $V_{a_k}$ in a closed
embedded sphere $S^{(k)}_L \cong \nolinebreak S^{\lambda _k
  -1}$. Since
dim$(S^{(k)}_L) = \lambda_{k-1} < n-1$, the open subset 
$\dot{D}_k\cap(V_{a_k}-S^{(k)}_L)$ is not empty in $V_{a_k}$.
\footnote
{The absence of
index-n critical points is a necessary condition. Consider
 an $\mathsf{N}$-shaped hollow cylinder. There is no
curve along which the height function increases monotonically connecting the initial
and final circles: it contains $\lambda = 2$ critical points.}
 Hence 
we can find a closed disk $D'_k \subset\dot{D}_k\cap(V_{a_k}-S^{(k)}_L)$. 
Projecting $D'_k$ backwards along the integral curves of $\xi$
gives a cylinder $C'_k \subset C_k$; projecting it forwards generates
a cylinder $C''_k$ from $V_{a_k}$ to $V_{a_{k+1}}$ that ends on what
we define to be $D_{k+1}\subset V_{a_{k+1}}$ and, by construction, does not intercept $p_k$. 
The union $C'_k \cup C''_k$ is the cylinder $C_{k+1}$ between 
$V_0$ and $V_{a_{k+1}}$ that contains no critical points, see
Fig \ref{cyl}.

This procedure is repeated for each elementary cobordism $f^{-1}[a_k,
a_{k+1}]$ until $V_{a_{r+1}}\equiv V_1$ is reached. The final disk
$D_{r+1}$ has finite size for the number of inductive steps is finite. The 
cylinder $C_{r+1}$ is the desired $C$. We select the integral curve 
$\mathbf{C}\equiv \{ p \in C :  \Psi (p)=(\vec{0}, t)\quad t\in[0,1]\}$, where 
the diffeomorphism $\Psi :C\rightarrow B^{n-1} \times B^1$ is a parametrisation
of the cylinder adapted to $\xi$ in the sense that $t(p)=f(p)$.

         That $L=M-\mathbf{C} $ is a cobordism between $U_0$ and
        $U_1$ follows from the equivalence under diffeomorphism
of all embedded $B^k$ balls
        in any connected manifold of dimension $n\geq k$
        \cite{palais59}; so in particular there is a diffeomorphism
       of $V_i$ that maps $q_i$ to $\tilde q_i$, $i = 0,1$. This
       implies that $V_i-q_i \cong V_i - \tilde q_i$ and so there
       exists a diffeo $d_i: V_i - q_i \rightarrow U_i$.
     To define a Morse metric in the
        manifold $L=M-\mathbf{C} $ we can combine local metrics as was
        done 
in the proof of lemma \ref{lemmametric}, even though $L$ is no longer 
        compact. This
        is because partitions of unity exist for the more general
        class of paracompact manifolds, and by construction, $L$ is
        certainly one of these \cite{dubrovin95}. But there are now two
        conditions to be simultaneously fulfilled: correct
        restrictions and asymptotic flatness. The former requires
        a sectioning of $L$ as in \ref{lemmadec1}; while
        the latter is achieved through a sectioning by
        cylinders concentric on asymptotic infinity. Note that, since
       $L$ is a submanifold of $M$, the restrictions $f_{|L}$ and 
$\xi_{|L}$  are a Morse function and a gradient-like vector field in
$L$. So using again small numbers 
$\delta < \epsilon$ with $\epsilon < c_1$ and  $c_r <1-\epsilon$
we cover $L$ with three open regions: the collarings $L_0, \;L_1$ 
        of the initial and final boundaries and the inner submanifold $L_2$.
\[\begin{array}{lll}
L_0\equiv & f^{-1}([\;0\;, \;\epsilon\;))\cap L
      &\stackrel{\Phi _0}{\cong}(V_0-q_0)\times [0, \epsilon)\\
L_1 \equiv & f^{-1}((1-\epsilon , 1]) \cap L 
       & \stackrel{\Phi _1}\cong (V_1-q_1)\times (1-\epsilon , 1] 
\end{array}\]
\[
L_2\equiv\; \; f^{-1} \left(\delta, 1-\delta )\right)\cap L  \hspace{4.33cm}
\]

Then we carry out the second sectioning: take two solid cylinders 
$C'$ and $C''$ concentric 
to $C$ in $M$, namely $\mathbf{C}\subset C'' \subset C' \subset C$. 
Define the core $L_{\mathrm{I}}\equiv L- C''$ and the asymptotic region 
$L_{\mathrm{II}}\equiv \dot{C}'-\mathbf{C}$ so that 
$L = L_{\mathrm{I}} \cup L_{\mathrm{II}}$. 
Choose a locally finite atlas 
$\mathcal{B}= (W_\alpha, \varphi_{\alpha})$ of
$L$ and define a refinement of $\mathcal{B}$ adapted to the
double sectioning. We obtain new charts $W_{\alpha i a }\equiv 
W_{\alpha}\cap L_i \cap L_a\quad i =0,1,2 \quad a=\mathrm{I}, \mathrm{II}$.  

Endow $V_0-q_0$ with the induced metric 
$\tilde h_0\equiv d_0^{\displaystyle *}(h_0)$ and let $H_0$ be the 
product Riemannian metric $ds^2=dt^2 +\tilde{h}_{0\, ab}\, dx^adx^b$ 
on $(V_0-q_0)\times [0,\epsilon)$. Define the metric $G_0\equiv \Phi _0
^{\displaystyle *} H_0$ on $L_0$. Similarly define the metric $G_1$ on $L_1$.
By asymptotic flatness of $\tilde h_0$ and $\tilde h_1$, 
$G_0$ and $G_1$ are on 
the right track for overall asymptotic flatness on $L$. The same asymptotic 
behaviour must hold throughout $L_{\mathrm{II}}$, {\it i.e.}, the metric must 
become flat in the surroundings of the removed curve. This is ensured by
taking $G_{\mathrm{II}}$ to be the pullback of $ds^2=dt^2 +dr^2 + r^2d\Omega ^2_{n-2}$ 
on $B^1\times (1, \infty)\times S^{n-2}\cong L_{\mathrm{II}}$ where
the $t$ coordinate is identified with $f$ via the diffeo.
 The local metrics on the
charts of the refinement are:
\[
\mbox{~~~~~~~~~~~~~~~~}G_{\mu \nu}^{\alpha i a}  =
\left\{\begin{array}{ll} (G_{i\, |W_{\alpha i a}})_{\mu \nu}  &\mbox{if $i=0,1\quad \forall \alpha, a$} \\
                           (G_{\mathrm{II}\, |W_{\alpha 2 \mathrm{II}}}) _{\mu \nu } & \forall \alpha  
        \end{array} \right.
\]
and any Riemannian metric $G_{\alpha 2 \mathrm{I}}$ on the 
remaining charts $W_{\alpha 2 \mathrm{I}}$. Finally, combine all these local 
metrics through a partition of unity $\{\vartheta_{\alpha i a}\}$ associated with the
atlas $(W_{\alpha i a}, \varphi_{\alpha |W_{\alpha i a}})$ 
to define a Riemannian metric
\[
        G(p) = \sum_{\alpha, i, a} \vartheta_{\alpha i a}(p)\;G^{\alpha i a}(p)
\]
which: (i) is asymptotically flat, (ii) has the correct restrictions
$\tilde h_0$ and $\tilde h_1$ 
and (iii) ensures $G^{\mu\nu }\partial_\mu f \partial_\nu f = 1$
 throughout $L_0 - L_0\cap L_2$ and $L_1 - L_1\cap L_2$. 
The Lorentzian metric formed from $G$ and $f$ as in equation (2)
inherits from $G$ properties (i) and (ii) and 
has the Morse structure
of $f$. Thus $g$ is the metric that we are seeking. $\Box $

        {\bf Proof of lemma \ref{lemmadec2}.} (i) Along the lines of the proof 
        just given, we will find an annular
        tube $A \cong B^1\times S^1\times B^{n-2}$ with central
        $\mathbf{A} \cong B^1\times S^1\times \{\vec{0}\}$ such that 
$V_i - \tilde C_i \cong V_i - C_i$, $i=0,1$, where $C_i = V_i \cap 
\mathbf{A}$.

        With the same notation as in the proof of lemma
        \ref{lemmadec1}, 
induction
        starts: thicken the initial circle $\tilde{C_0}$ to an
        embedded ``torus"\footnote{The existence of tubular
        neighbourhoods of submanifolds is a well established fact in
        differential topology.}, $i_0: S^1\times B^{n-2}\rightarrow T_0\subset V_0$ and
        project it along the $\xi$-lines to $V_{a_1}\cong V_0$. This
        gives a first annular tube $A_1$. Now suppose that an $A_k$
        has been found from $V_0$ to $V_{a_k}$ that satisfies:
\begin{enumerate}
\item $A_k \cap V_0 = i_0 (S^1\times B_{x_k})$, where $B_{x_k} \subset
\dot{B}^{n-2}$ is a ball centred at a point $x_k \in B^{n-2}$ which is
not necessarily the origin.

\item $A_k \cap V_{a_k}\equiv T_k =i_k(S^1\times B_{x_k})$ with
$i_k(\psi, x)\equiv \sigma^{\xi}_{i_0(\psi, x)}\cap
V_{a_k}$ for each $\psi\in [0,2\pi ),x\in B_{x_k}$. Here $\sigma ^{\xi}_q$ denotes 
the image in $M$ of the integral curve starting at $q\in V_0$.

\end{enumerate}

We find an annulus $A_{k+1}$ of integral curves of $\xi$ in
$M^{a_{k+1}} \equiv f^{-1}([0,a_{k+1}])$ 
whose intersection with $V_0$ is still a thickened
circle of the form $i_0 (S^1\times B_x)$.  Recall that between
$V_{a_k}$ and $V_{a_{k+1}}$ there is a Morse point $p_k$ whose index
is now constrained by $\lambda_k \leq n-2$. So $\dot{T}_k \cap
(V_{a_k} -S^{(k)}_L)$ is not empty and since $S^{(k)}_L$ has dimension
less or equal to $n-3$ we can in fact choose another $T'_k\subset
\dot{T}_k$ such that $T'_k=i_k(S^1\times B_{x_{k+1}})$ for some
(n-2)-ball $B_{x_{k+1}} \subset B_{x_k}$ and that $T'_k \cap S^{(k)}_L
=\emptyset $. Projecting $ T'_k$ backwards along the integral curves
of $\xi$ gives an annular tube $A'_k \subset A_k$; projecting it
forwards generates another tube $A''_k$ from $V_{a_k}$ to
$V_{a_{k+1}}$ that ends on what we denote $T_{k+1}$ and does not
intercept $p_k$. The union $A'_k \cup A''_k$ is the desired thickened
cylinder $A_{k+1}$ between $V_0$ and $V_{a_{k+1}}$ that contains no
critical points.

Again, apply the inductive step to each elementary cobordism
 until $V_{a_{r+1}} \equiv V_1$ is reached. Like
before, the thickened circle $T_{r+1}$ has finite size. Now 
$A_{r+1}$ is the annular tube $A$ between $V_0$ and $V_1$, 
which can be parametrised as the product $B^1\times B^{n-2} \times S^1$: for $p\in A$ the coordinates are 
$(f(p), i_0(q))$,
where $q$ is the point of $T_0$ at which starts the $\xi$-curve
through $p$ and the second coordinate naturally splits in an $S^1$ and
a $B^{n-2}$ parts. We select the integral annulus $\mathbf{A}\equiv \{p
\in C : q = i_0(\psi, x_{r+1})\;0\leq \psi <2\pi \}=
\{\sigma^{\xi}_{i_0(\psi, x_{r+1})} (t) : 0\leq \psi <2\pi , 0 \leq t
\leq 1\}$. The subsets $C_0\equiv \mathbf{A} \cap V_0$ and 
$C_1\equiv \mathbf{A} \cap V_1$ are embedded circles in the initial 
and final boundaries.

 To demonstrate that $L\equiv M-\mathbf{A}$ is a cobordism
        between $U_0$ and $U_1$ we need to show that $V_0 - C_0 \cong
        V_0
- \tilde C_0$ and $V_1 - C_1 \cong V_1 - \tilde C_1$. For the former,
consider
that $\tilde C_0$ is the image under $i_0$ of $\{(\psi, \vec{0}): \psi
\in S^1\}$ in $V_0$ and $C_0$ is the image under $i_0$ of 
$\{(\psi, {\vec{x}}_{r+1}): \psi
\in S^1\}$. The result by Palais shows that there is a diffeo of $S^1 \times
B^{n-2}$ that maps $(\psi, \vec{0})$ to $(\psi, {\vec{x}}_{r+1})$ and 
is the identity on the boundary. Then $i_0$ turns this into a diffeo 
of $V_0$ that maps $\tilde C_0$ onto $C_0$ and hence the result. 
For the latter, recall that $V_1$ is simply connected and connected
by assumption and dim($V_1$)$\ge 4$. These conditions guarantee the absence of
knots and also imply that every embedded circle bounds a disk; in particular 
$\tilde C_1$ and $C_1$ do. There exists a diffeo of $V_1$ that maps one of these 
disks onto the other and so maps $\tilde C_1$ onto $C_1$. Hence the result.

        Finally the asymptotically flat AL metric in $L$ is defined as
        in the proof of \ref{lemmadec1}; the only difference resides in the
        sectioning of manifold $L$: to
        perform the transition between an arbitrary Riemannian metric
        in the core of $L$ and the flat metric around the annular
        cylinder at infinity $\mathbf{A}$, we use concentric
        annular tubes instead of concentric cylinders. $\Box $

\end{document}